\documentclass[reqno, 11pt]{amsart}
\usepackage{enumerate}
\newtheorem{Theorem}{Theorem}
\newtheorem{Proposition}{Proposition}
\newtheorem{Lemma}{Lemma}
\newtheorem{Definition}{Definition}

\numberwithin{equation}{section}

\newcommand{\vo}{(1 + g_{rs} v^r v^s)}

\newcommand{\EoPr}{\hfill $\Box$}
\newcommand{\Proof}{{\sc Proof : }}

\title{The Einstein-Vlasov system with a scalar field}
\author{Hayoung Lee}
\address{Max-Planck-Institut f\"ur Gravitationsphysik,
Am M\"uhlenberg 1, Golm bei Potsdam, D-14476, Germany}
\email{hayoung@aei.mpg.de}

\begin{document}

\begin{abstract}
We study the Einstein-Vlasov system coupled to a nonlinear scalar field
with a nonnegative potential in locally spatially 
homogeneous spacetime, as an expanding cosmological model.
It is shown that solutions of this system exist globally in time.
When the potential of the scalar field is of an exponential form,
the cosmological model corresponds to accelerated expansion.
The Einstein-Vlasov system coupled to a nonlinear scalar field whose potential is
of an exponential form  
shows the causal geodesic completeness of the spacetime towards the future.
The asymptotic behaviour of solutions of this system
in the future time is analysed in various aspects, 
which shows power-law expansion.
\end{abstract}
\maketitle

\section{Introduction}

Particle systems are modeled statistically by
distribution functions, which at any time represent the probability
to find a particle in a given position, with a given momentum.
The distribution functions contain a wealth of information and macroscopic quantities are
calculated from these functions.
The models being considered here are those in which collisions between particles are sufficiently rare
to be neglected. The collection of these collisionless particles
is described by Vlasov equations. For this reason,
matter considered in these physical models is said to be collisionless matter
or Vlasov matter.

The time evolutions of particle systems are determined by the interactions between the particles
which rely on the physical situation.
Each particle is driven by self-induced fields which are generated by all particles together.
Naturally combinations of interaction processes are also considered 
but in many situations, one of them is strongly dominating and the weaker processes are neglected.
In gravitational physics, these fields are described by the Einstein equations.
The physical models concerned in this paper is described by
the Vlasov equation which is coupled to the Einstein equations
by means of the energy-momentum tensor. One application of the Vlasov
equation coupled to this self-gravitating system is cosmology.
The particles are in this case galaxies or even clusters of galaxies.

The simplest cosmological models are those which are spatially homogeneous.
Spatially homogeneous spacetimes can be classified into two types;
Bianchi models and the Kantowski-Sachs models.
The models with a three-dimensional group of isometries $G_3$ acting simply
transitively on spacelike hypersurfaces are Bianchi models. There are nine types I-IX,
depending on the classification of the structure constants of the Lie algebra of $G_3$.
Those admitting a group of isometries $G_4$ which acts on spacelike hypersurfaces but no subgroup
$G_3$ which acts transitively on the hypersurface are Kantowski-Sachs models.
In fact, $G_3$ subgroup acts multiply transitively on two-dimensional spherically symmetric surfaces.

If we take as a cosmological spacetime one which admits a compact Cauchy hypersurface,
the Bianchi types which can occur for a spatially homogeneous cosmological model are
only type I and IX and also Kantowski-Sachs models.
Because of the existence of {\em locally} spatially homogeneous cosmologies,
we take a larger class of spacetimes possessing a compact Cauchy hypersurface
so that this allows a much bigger class of Bianchi types to be included.
Since the Cauchy problem for the Einstein-Vlasov system is well-posed,
it is enough to define the class of initial data.
Here is the definition.
\begin{Definition}
{\rm Let $\stackrel{\rm _o}{g}_{ij}$, $\stackrel{\rm _o}{k}_{ij}$ and
$\stackrel{\rm _o}{\mathcal{F}}$ be initial data for a Riemannian metric,
a second fundamental form, and a matter, respectively,
on a three-dimensional manifold $M$.
Then this initial data set $(\stackrel{\rm _o}{g}_{ij}, \stackrel{\rm _o}{k}_{ij}, \stackrel{\rm _o}{\mathcal{F}})$
for the Einstein-Vlasov system is called {\em locally spatially homogeneous}
if the naturally associated data set on the universal covering $\widetilde{M}$ is
homogeneous, i.e., invariant under a transitive group action.}
\end{Definition}
So the spacetimes considered here will be Cauchy developments
of locally homogeneous initial data sets on some manifolds.
Note that a complete Riemannian manifold is
locally homogeneous if and only if the universal cover is homogeneous.
For Bianchi models the universal covering space can be identified with a Lie group $G$.
So the natural choice for $G$ in this case is a simply connected three-dimensional Lie group.
(For a detailed discussion on this subject we refer to \cite{Ren1, Ren2}).

In this paper, we discuss the dynamics of expanding cosmological models,
particularly accelerated expansion. There are two subjects concerning
this rapid expansion. One is the very early universe close to the big bang (inflation)
and the other is the present era (quintessence)
supported by the observations of supernovae of type Ia.

One simple way to obtain accelerated expansion is
to introduce a positive cosmological constant, which leads to exponential expansion.
In homogeneous spacetimes it has been studied by Wald in \cite{Wald1} with general matter
which satisfies the dominant and strong energy conditions.
When the matter is described by the Vlasov equation,
the detailed asymptotics of solutions have been analysed in \cite{Lee}.
In the inhomogeneous case Vlasov matter model has been studied in \cite{Tch,TchRen}
under some symmetric conditions.
In \cite{Ren2003} by Rendall, vacuum and perfect fluid cases are handled.

Another choice for accelerated expanding cosmological models, which is more
sophisticated, is a nonlinear scalar field. 
It has been analysed by Rendall in \cite{Ren2004}
that when the potential of the scalar field has a positive lower bound
with general matter satisfying the dominant and strong energy conditions then
the homogeneous models expand exponentially.
In the case of an exponential potential, the models shows power-law expansion
which has been studied in \cite{KM1,KM2} 
by Kitada and Maeda.

Bianchi type IX and Kantowski-Sachs models have complicated features
when a positive cosmological constant or a nonlinear scalar field is present.
It has been seen that
there are chaotic behaviours between expanding and recollapsing phases in these models.
In the discussion of expanding cosmology,
the models being concerned in this paper 
are all Bianchi types except IX, with a nonlinear scalar field and
vanishing cosmological constant.

The fields described by the Einstein equations are coupled to
the Vlasov equation and a nonlinear scalar field by the energy-momentum tensor
which is of form
\begin{equation}\label{tensor}
T_{\alpha\beta}= T_{\alpha\beta}^{(Vlasov)} + \nabla_\alpha \phi \nabla_\beta \phi
- \big(\frac{1}{2}\nabla^\gamma\phi\nabla_\gamma\phi + V(\phi)\big) g_{\alpha\beta}
\end{equation}
where $\phi$ is a scalar field  which represents dark energy,
V is a potential and
$T_{\alpha\beta}^{(Vlasov)}$ is the energy-momentum tensor of the collisionless matter
described by the Vlasov equation. 
$T_{\alpha\beta}^{(Vlasov)}$ satisfies
the dominant and strong energy conditions given respectively by
\begin{enumerate}[(1)]
\item $T_{\alpha\beta}^{(Vlasov)} v^\alpha w^\beta \geq 0$
where $v^\alpha$ and $w^\beta$ are any two future pointing timelike vectors,
\item $\big(T_{\alpha\beta}^{(Vlasov)}
- \frac{1}{2}g_{\alpha\beta}(T_{\mu\nu}^{(Vlasov)} g^{\mu\nu})\big)
v^\alpha v^\beta \geq 0$
for any timelike vector $v^\alpha$
\end{enumerate}
As a consequence of the Bianchi identity in (\ref{tensor}) the scalar field $\phi$
satisfies the equation
$$
\nabla_\alpha \nabla^\alpha \phi = V'(\phi)
$$
To make the notation not too heavy the superscript $(Vlasov)$ will be omitted
for the rest of the paper.

The content of the rest of this paper is the following.
Section \ref{sec:form} presents the detailed formulation of the system being considered.
In Section \ref{exist.geodesic},
we prove the global existence of solutions for the Einstein-Vlasov system
coupled to a nonlinear scalar field with a general potential.
Also the causal geodesic completeness
of the spacetime towards the future will be present, in the case of an exponential potential.
In Section \ref{asymp},
we study the asymptotic behaviour of solutions at late times in various aspects,
when the potential of the scalar field is of an exponential form.
We observe the future asymptotic behaviours of the mean curvature, the metric,
the momenta of particles along the characteristic curves as well as the generalized Kasner
exponents and the deceleration parameter.
Also we analyse the energy-momentum tensor in an orthonormal frame on the hypersurfaces.
As we will see later on, 
the cosmological model being considered in this paper exhibits power-law expansion.

\section{Einstein-Vlasov system with a scalar field}\label{sec:form}
Here is the formulation of the Einstein-Vlasov system coupled to a nonlinear scalar field
with a potential.
Let $G$ be a simply connected three-dimensional Lie group and $\{e_i\}$, 
a left invariant frame and $\{e^i\}$, the dual coframe.
Consider the spacetime as a manifold $G \times I$, where $I$ is an open interval and 
the spacetime metric of our model has the form
\begin{equation}\label{metric}
ds^2 = -dt^2 + g_{ij}(t) e^{i} \otimes e^{j}
\end{equation}
The initial value problem for the Einstein-Vlasov system is investigated in the case of this special form 
of the metric and the distribution function $f$ depends only on $t$ and $v^i$, 
where $v^i$ are spatial components of the momentum in the frame ${e_i}$. 
Initial data will be given on the hypersurface $G \times \{t_0\}$.

Now the constraints are
\begin{gather}
R-(k_{ij}k^{ij}) + (k_{ij}g^{ij})^2 = 16 \pi T_{00} + 8\pi\psi^2 + 16 \pi V(\phi) \label{cons1}\\
\nabla^i k_{ij} = - 8 \pi T_{0j} \label{cons2}
\end{gather}
The evolution equations are
\begin{align}
\frac{d}{dt}g_{ij} & = -2 k_{ij} \label{evol1}\\
\frac{d}{dt}k_{ij} & = R_{ij} + (k_{lm}g^{lm}) k_{ij} - 2 k_{il} k^l_j - 8 \pi T_{ij} \label{evol2}\\
		  &\qquad  - 4\pi T_{00} g_{ij} + 4 \pi (T_{lm}g^{lm}) g_{ij} - 8 \pi V(\phi) g_{ij} \notag\\
\frac{d}{dt}\phi &= \psi \label{evol3}\\
\frac{d}{dt}\psi &= (k_{lm}g^{lm}) \psi - V'(\phi) \label{evol4}
\end{align}
These equations are written using frame components. Here $k_{ij}$ is
the second fundamental form, $R$ is the Ricci scalar curvature and $R_{ij}$ is the Ricci
tensor of the three-dimensional metric. 
And $\phi$ is a scalar field, depending only on $t$, with a nonnegative potential $V(\phi)$.
$\psi$ is a function introduced by the relation (\ref{evol3}).
The nonnegative assumption on the potential is very natural.
It implies that the dominant energy condition is satisfied and then 
the weak energy condition follows.

Here are the components of the energy-momentum tensor of the Vlasov matter ;
\begin{align}
T_{00} (t) & = \int f(t, v) \vo^{1/2} (\det g)^{1/2} \,dv\label{tensor1}\\
T_{0i} (t) & = \int f(t, v) v_i (\det g)^{1/2} \,dv\label{tensor2}\\
T_{ij} (t) & = \int f(t, v) v_i v_j \vo^{-1/2} (\det g)^{1/2} \,dv\label{tensor3}
\end{align}
Here $v:=(v^1, v^2, v^3)$ and $dv :=dv^1\,dv^2\,dv^3$.

The Vlasov equation is
\begin{equation} \label{vlasov}
\partial_t f + 
  \big\{ 2 k^i_j v^j - \vo^{-1/2}\gamma^i_{mn} v^m v^n \big\} \partial_{v^i}f = 0
\end{equation}
Here the Ricci rotation coefficients $\gamma^i_{mn}$ are defined as
$$
\gamma^i_{mn} = \frac{1}{2} g^{ik} (-C^l_{nk} g_{ml} + C^l_{km} g_{nl} + C^l_{mn} g_{kl})
$$
where $C^i_{jk}$ are the structure constants of the Lie algebra of $G$.
To have a complete set of equations it is necessary to compute $R_{ij}$ in terms of
$g_{ij}$. In this paper, it is enough to know that $R_{ij}$ is of the form
$(\det g)^{-n}$(polynomial in $g_{ij}$ and $C^i_{jk}$).
To control $(\det g)^{-1}$ we use 
\begin{equation}\label{eqn:log_g}
\frac{d}{dt} \log (\det g) = - 2 (k_{ij} g^{ij})
\end{equation}

Note that in discussing expanding cosmological models, 
the sign convention for $(k_{ij}g^{ij})$ in the paper is negative.
Also it is true that if models are initially expanding, i.e., $(k_{ij}g^{ij})(t_0) < 0$
then $(k_{ij}g^{ij})(t) <0$ for all time $t \geq t_0$ (For details see \cite{Ren2004}).

The evolution equations are in general partial differential equations, i.e.
$d/dt$ is $\partial_t$. However due to the locally
spatially homogeneous spacetime, the partial differential equations are reduced to 
ordinary differential equations.

For the rest of the paper, $C$ denotes a positive constant which
changes from line to line and may depend only on the initial data.
Also $C_l$ ($l=0, 1, 2, \ldots$) are positive constants.

\section{Global existence of solutions and geodesic completeness}\label{exist.geodesic}
In this section, we will show global existence solutions of
the Einstein-Vlasov System coupled to a nonlinear scalar field
with a potential.
As a first step, conditions will be established under which solutions of this system
exist globally in time, with the technique appeared in \cite{Ren1}
by which the existence of solutions for the Einstein-Vlasov system 
in the absence of a scalar field has been proved.
And then eventually it will be proved that these conditions are fulfilled
in the system being considered.
Also we will observe the casual geodesic completeness of the spacetime towards the future direction,
when the potential of the scalar field is of an exponential form. 
\begin{Proposition}\label{prop:GlobalExis}
Let $g_{ij}(t_0)$, $k_{ij}(t_0)$, $\phi(t_0)$, $\psi(t_0)$ and $f(t_0, v)$ be 
an initial data set for the evolution equations (\ref{evol1}) -- (\ref{evol4}) 
and the Vlasov equation (\ref{vlasov}) which has Bianchi symmetry and 
satisfies the constraints (\ref{cons1}) and (\ref{cons2}).
Also let $f(t_0, v)$ be a nonnegative $C^1$ function with compact support.
And assume that the potential of the scalar field $V(\phi)$ is a nonnegative $C^2$ function. 
Then there exists a unique $C^1$ solution $(g_{ij},\, k_{ij},\, \phi,\, \psi,\, f)$ 
of the Einstein-Vlasov system, on an interval $[t_0, T)$, for some time $T$. 
If $|g|$, $(\det g)^{-1}$, $|k|$, $|\phi|$, $|\psi|$, $\|f\|$ 
and the diameter of supp $f$ are bounded on $[t_0, T)$, then $T=\infty.$
\end{Proposition}
\Proof
The characteristics of (\ref{vlasov}) are the solutions $V^i(s, t, v)$ of the equation
 \begin{equation}\label{characteristics}
 \frac{dV^i}{ds} = 2 k^i_j V^j - (1+g_{rs}V^rV^s)^{-1/2}\gamma^i_{mn} V^m V^n
 \end{equation}
with $V^i(t, t, v)=v^i$.
Let $f^{(0)} (t, v) = f(t_0, v)$, $g_{ij}^{(0)}(t) = g_{ij}(t_0)$, $k_{ij}^{(0)}(t) = k_{ij}(t_0)$,
$\phi^{(0)} (t) = \phi(t_0)$ and $\psi^{(0)} (t) = \psi(t_0)$.
If $f^{(n)}$, $g_{ij}^{(n)}$, $k_{ij}^{(n)}$, $\phi^{(n)}$ and $\psi^{(n)}$ are given for some $n$,
determine $V^{(n+1)}$ by solving the characteristic equation (\ref{characteristics})
with $k_{ij}^{(n)}$ and $g_{ij}^{(n)}$. Let $f^{(n+1)}(t, v) = f(t_0, V^{(n+1)}(t_0, t, v))$.
Define an energy-momentum tensor $T_{\alpha\beta}^{(n+1)}$ with $f^{(n+1)}$ and $g_{ij}^{(n)}$
in (\ref{tensor1}) -- (\ref{tensor3}). Determine $g_{ij}^{(n+1)}$, $k_{ij}^{(n+1)}$, $\phi^{(n+1)}$ and $\psi^{(n+1)}$
by solving (\ref{evol1}) -- (\ref{evol4}) 
with $T_{\alpha\beta}^{(n+1)}$, $g_{ij}^{(n)}$, $k_{ij}^{(n)}$, $\phi^{(n)}$ and $\psi^{(n)}$
in the right hand side of equations and with $g_{ij}^{(n+1)}$, $k_{ij}^{(n+1)}$, $\phi^{(n+1)}$
and $\psi^{(n+1)}$ in the left hand side. 
Now let $[t_0, T^{(n+1)})$ be the maximal interval on which $g_{ij}^{(n+1)}$ is positive definite.
By induction, one can see that $f^{(n)}$, $g_{ij}^{(n)}$, $k_{ij}^{(n)}$, $\phi^{(n)}$ and $\psi^{(n)}$ 
are $C^1$ on their domains of definition.

Let $|g|$ be the maximum modulus of any component $g_{ij}$ and $|k|$ for $k_{ij}$.
Suppose that for all $n \leq N-1$ the following bounds hold:
\begin{gather}\label{ggk}
|g^{(n)}-g^{(0)}| \leq A_1, \quad (\det g^{(n)})^{-1} \leq A_2, \quad |k^{(n)}-k^{(0)}|\leq A_3\\
|\phi^{(n)}-\phi^{(0)}| \leq A_4, \quad |\psi^{(n)}-\psi^{(0)}| \leq A_5 
\end{gather}
Also suppose that $|v| \leq A_6$ whenever $f^{(n)}(t, v) \neq 0$. Here $A_i$ ($i=1,\ldots,6$) 
are positive constants which are for the moment arbitrary. The characteristic system (\ref{characteristics})
implies a bound for the form
$|v| \leq C_0 + B_6 t$ whenever $f^{(N)}(t, v)\neq 0$, where $B_6$ depends only on $A_i$'s.
As a consequence (\ref{tensor1}) -- (\ref{tensor3}) imply a bound for $T_{\alpha\beta}^{(N)}$
depending only on $A_i$'s. The evolution equations (\ref{evol1}) -- (\ref{evol4})
imply bounds of the form
\begin{gather*}
|g^{(N)}-g^{(0)}| \leq B_1t, \quad |k^{(N)}-k^{(0)}|\leq B_3t \\
|\phi^{(N)}-\phi^{(0)}| \leq B_4 t, \quad|\psi^{(N)}-\psi^{(0)}| \leq B_5 t
\end{gather*} 
where $B_i$'s depend only on $A_i$'s.
If $A_i$'s are fixed then the inequalities in (\ref{ggk}) imply an inequality of the form
$(\det g^{(N)})^{-1} \leq B_2$ whenever $t \leq T$ and $T$ is some positive time depending only on $A_i$'s.
Now fix $A_i$'s in such a way that $A_2 > (\det g^{(0)})^{-1}$ and $A_6 > C_0$.
Next reduce the size of $T$ if necessary so that $B_iT <A_i$ (i=1, 3, 4, 5), 
$B_2 < A_2$ and $C_0 +B_6T <A_6$.
Then all iterates exist on the interval $[t_0, T)$ 
and $g^{(n)}$, $k^{(n)}$, $\phi^{(n)}$ and $\psi^{(n)}$ are bounded on that interval independently of $n$.

Now we need to show that these iterations converge.
Consider the difference of successive iterates for $n \geq 2$,
\begin{align}
&|(g^{(n+1)}-g^{(n)})(t)|+|(k^{(n+1)}-k^{(n)})(t)|\notag\\
&\qquad +|(\phi^{(n+1)}-\phi^{(n)})(t)|+|(\psi^{(n+1)}-\psi^{(n)})(t)|\notag\\
& \quad \leq  C \int^t_{t_0} \left[|(g^{(n)}-g^{(n-1)})(s)|+|(k^{(n)}-k^{(n-1)})(s)|
  			  +|(\phi^{(n)}-\phi^{(n-1)})(s)|\right.\notag\\
& \qquad \left. + |(\psi^{(n)}-\psi^{(n-1)})(s)| + \|(f^{(n+1)}-f^{(n)})(s)\|_\infty \right] \, ds\label{eqn:iter}
\end{align}
For the difference of the characteristics note that
\begin{equation}\label{eqn:diffV}
\Big|\frac{d}{ds}V^{(n+1)}-\frac{d}{ds}V^{(n)}\Big| 
\leq C\big[|V^{(n+1)}-V^{(n)}|+|g^{(n)}-g^{(n-1)}|+|k^{(n)}-k^{(n-1)}|\big]
\end{equation}
Define
\begin{align}\label{def:alpha}
&\alpha^{(n)}(t):= |(g^{(n+1)}-g^{(n)})(t)| + |(k^{(n+1)}-k^{(n)})(t)|\\
&\;+ |(\phi^{(n+1)}-\phi^{(n)})(t)| + |(\psi^{(n+1)}-\psi^{(n)})(t)|\notag\\
&\;+ \sup \{|V^{(n+1)}-V^{(n)}|(s, t, v) : s \in [t_0, t],\, 
			   	  v \in {\rm supp} f^{(n+1)}(t) \cup {\rm supp} f^{(n)}(t)\}\notag
\end{align}
Then we get
\begin{equation}\label{eqn:diff_f}
\|f^{(n+1)}(t)-f^{(n)}(t)\|_\infty \leq \|f^{(0)}\|_{C^1} \alpha^{(n)}(t)
\end{equation}
Therefore (\ref{eqn:iter}) -- (\ref{eqn:diff_f}) imply that
$$
\alpha^{(n)}(t) \leq C \int^t_{t_0} \big[ \alpha^{(n)}(s) + \alpha^{(n-1)}(s)\big] \,ds
$$
Applying Gr\"onwall's inequality to this gives 
$$
\alpha^{(n)}(t) \leq C \int^t_{t_0} \alpha^{(n-1)}(s) \,ds
$$
Therefore $\alpha^{(n)}(t) \leq C^{n-2} \|\alpha^{(2)}\| t^{n-2}/(n-2)!$
and so $\{g^{(n)}\}$, $\{k^{(n)}\}$, $\{\phi^{(n)}\}$, $\{\psi^{(n)}\}$ and $\{V^{(n)}\}$
are Cauchy sequences on the time interval $[t_0, T)$. Denote the limits of these sequences by
$g$, $k$, $\phi$, $\psi$ and $V_\infty$, respectively. Also by (\ref{evol1}) -- (\ref{evol4}),
$dg^{(n)}/dt$, $dk^{(n)}/dt$, $d\phi^{(n)}/dt$, $d\psi^{(n)}/dt$
and $dV^{(n)}/dt$ are uniformly convergent. 
Thus $(g, k, \phi, \psi, f)$ is a $C^1$ solution of the system on the interval $[t_0, T)$.

Now let us check whether a solution exists uniquely or not. 
If two solutions with the same initial data are given, define a quantity $\alpha(t)$ in terms
of their difference in the same way that $\alpha^{(n)}(t)$ was defined
in terms of the differences of two iterates. Applying the same argument as above leads to
an estimate of the form 
$$\alpha(t) \leq C \int^t_{t_0} \alpha(s) \, ds$$
By Gr\"onwall's inequality we can see that $\alpha(t)$ is zero and hence that the two solutions agree.
Therefore the solution which has been constructed is uniquely determined by the initial data.   

Define
\begin{align*}
A &:= R-(k_{ij}k^{ij}) + (k_{ij}g^{ij})^2 - 16 \pi T_{00} - 8\pi\psi^2 - 16 \pi V(\phi) \\
A_i &:= \nabla^i k_{ij} + 8 \pi T_{0j}
\end{align*}
Then after a lengthy calculation we obtain
\begin{align*}
\frac{d}{dt} A &= 2(k_{ij}g^{ij})A -2 \gamma^l_{ij}g^{ij}A_l\\
\frac{d}{dt} A_i &= (k_{lm}g^{lm})A_i + 2 k^l_i A_l
\end{align*}
That is, (\ref{evol1}) -- (\ref{vlasov}) imply a homogeneous first
order ordinary differential system for constraints (\ref{cons1}) and (\ref{cons2}).
Therefore we can conclude that if the initial data satisfy the constraints 
then so does the solution of the evolution equations (\ref{evol1}) -- (\ref{evol4}) 
with energy-momentum tensors (\ref{tensor1}) -- (\ref{tensor3}) and the Vlasov equation
(\ref{vlasov}).

In the above argument so far, we see that the size of $T$ is only restricted by the quantities ;
$|g^{(0)}|$, $(\det g^{(0)})^{-1}$, $|k^{(0)}|$, $|\phi^{(0)}|$, $|\psi^{(0)}|$, $\|f^{(0)}\|$
and the diameter of supp $f^{(0)}$. Thus if  
the quantities $|g|$, $(\det g)^{-1}$, $|k|$, $|\phi|$, $|\psi|$, $\|f\|$ and the diameter of supp $f$ are bounded
on the same time interval $[t_0, T)$,
then a solution exists on $[t, t+\epsilon)$ for any $t \in [t_0, T)$ and some $\epsilon$ independent of $t$.
It can be concluded that the original solution can be extended to the larger interval $[t_0, T+\epsilon)$. 
Therefore this completes the proof.
\EoPr

Let us state some properties of linear algebra which can be found in \cite{Lee, Ren1}.
We shall make use of these properties later on.
Let $A$ be a $n \times n$ matrix. 
Let $A_1$ and $A_2$ be $n \times n$ symmetric matrices with $A_1$ positive definite. 
Define {\em a norm} of a matrix by
\begin{equation*}
\|A\| := \sup \{\|Ax\|/\|x\| : x \neq 0,\, x \in \mathbb{R}^n\}
\end{equation*}
Also define {\em a relative norm} by 
\begin{equation*}
\| A_2 \|_{A_1} := \sup \{ \|A_2 x \| / \|A_1 x \| : x \neq 0 , \, x \in \mathbb{R}^n \}
\end{equation*}
Then from these definitions, one can see that
\begin{equation}\label{norm}
\|A_2\| \leq \|A_2\|_{A_1}\|A_1\|
\end{equation}
and also
\begin{equation}\label{relnorm}
\|A_2\|_{A_1} \leq \big( \text{tr}(A_1^{-1} A_2 A_1^{-1} A_2) \big)^{1/2}
\end{equation}

\begin{Proposition}\label{Prop:exist2}
If $(k_{ij}g^{ij})$ is bounded on $[t_0, T)$, then $T = \infty$.
\end{Proposition}
\Proof
Let $\sigma_{ij}$ be the trace free part of the second fundamental form $k_{ij}$.
Then we have $k_{ij} = \frac{1}{3}(k_{ij}g^{ij})g_{ij}+\sigma_{ij}$.
By this fact, we rewrite the constraint (\ref{cons1}) as
\begin{equation}\label{eqn:kg^2}
\frac{1}{3}(k_{ij}g^{ij})^2 = -\frac{1}{2}R + \frac{1}{2}(\sigma_{ij} \sigma^{ij})
				   + 8 \pi T_{00} + 4 \pi \psi^2 + 8 \pi V(\phi)
\end{equation}
It has been proved by Wald in \cite{Wald1} that in all Bianchi models except type IX, 
the Ricci scalar curvature is zero or negative. Also due to the nonnegative potential, we get
$$
\frac{1}{3}(k_{ij}g^{ij})^2 \geq 4 \pi \psi^2
$$
So if $(k_{ij}g^{ij})$ is bounded on $[t_0, T)$, then $\psi$ is bounded on $[t_0, T)$
and so is $\phi$.

From evolution equations (\ref{evol1}) and (\ref{evol2}), we have
\begin{equation}\label{eqn:kg_t}
\frac{d}{dt}(k_{ij}g^{ij}) 
	= R + (k_{ij}g^{ij})^2 + 4 \pi (T_{ij} g^{ij}) - 12\pi T_{00} - 24\pi V(\phi)
\end{equation}
Using the constraint (\ref{cons1}) we get
\begin{equation*}
\frac{d}{dt} (k_{ij}g^{ij}) = (k_{ij}k^{ij}) + 4\pi (T_{ij}g^{ij}) + 4\pi T_{00} 
+ 8 \pi \psi^2 - 8 \pi V(\phi)
\end{equation*}
Thus we obtain
\begin{equation*}
\frac{d}{dt}(k_{ij}g^{ij}) \geq (k_{ij}k^{ij}) - 8 \pi V(\phi)  
\end{equation*}
Then
$$
(k_{ij}g^{ij}) + 8 \pi \int^t_{t_0} V(\phi)(s) \, ds 
			   \geq k_{ij}(t_0)g^{ij}(t_0) + \int^t_{t_0} (k_{ij}k^{ij}) (s)\, ds
$$
Note that
$$
\int^t_{t_0} V(\phi)(s) \,ds 
\leq \|V'\|_t \int^t_{t_0}  |\phi(s)| \,ds + C(t+1)
$$
where $\|V'\|_t :=\sup \{|V'(\phi)(s)| :  \text{ for all } s \in [t_0, t] \}$.
Since $\phi$ is bounded on $[t_0, T)$, then $\int^t_{t_0} V(\phi)(s) \, ds$ is bounded for all $t$ in $[t_0, T)$.
Therefore with the boundedness of $(k_{ij}g^{ij})$, we conclude that
$\int^{T}_{t_0} (k_{ij}k^{ij}) (s)\, ds< \infty$.

Let $\|g\|$ and $\|k\|$ be the norms of the matrices with entries $g_{ij}$ and $k_{ij}$, respectively.
Let $\|k\|_g$ be the relative norm of the matrix with entries $k_{ij}$ with respect to the matrix
with entries $g_{ij}$.
Then using (\ref{relnorm}) we have
\begin{align*}
\|g(t)\| & \leq \|g(t_0)\| + 2 \int^t_{t_0} \|k(s)\| \,ds\\
		 & \leq \|g(t_0)\| + 2 \int^t_{t_0} \|k(s)\|_g \|g(s)\| \,ds\\
		 & \leq \|g(t_0)\| + 2 \int^t_{t_0} (k_{ij}k^{ij})^{1/2}(s) \|g(s)\| \,ds
\end{align*}
By Gr\"onwall's inequality, we get
$$
\|g(t)\| \leq \|g(t_0)\| \exp\left[2\int^t_{t_0} (k_{ij}k^{ij})^{1/2}(s) \,ds\right]
$$
Since $\int^t_{t_0} (k_{ij}k^{ij})^{1/2}(s)\,ds$ is bounded on $[t_0, T)$,
also $|g|$ is bounded on $[t_0, T)$.
Using (\ref{eqn:log_g}), we see that $(\det g)^{-1}$ is bounded on the same interval.
It is known that if $(\det g)$ and its inverse are bounded then the scalar curvature
$R$ is bounded from above.
Note that in (\ref{cons1}) we have
$$ 
R + (k_{ij}g^{ij})^2 \geq (k_{ij}k^{ij})
$$
Thus $k_{ij}k^{ij}$ is bounded on $[t_0, T)$.
By the inequality 
$$\|k\| \leq (k_{ij}k^{ij})^{1/2} \|g\|$$ 
also $|k|$ is bounded.
The boundedness of $|g|$ and $(\det g)^{-1}$ implies that
$g$ is uniformly positive definite on the interval. Hence the solutions
of the characteristic equation are also bounded.
Therefore by Proposition \ref{prop:GlobalExis}, the proof completes.
\EoPr

So far it has been proved that solutions of the system exist as long as 
some quantities are bounded in a finite time interval $[t_0, T)$ for arbitrary $T$. 
These conditions are satisfied, as we will see in the following theorem.
\begin{Theorem}\label{thm:exist}
Let $g_{ij}(t_0)$, $k_{ij}(t_0)$, $\phi(t_0)$, $\psi(t_0)$ and $f(t_0, v)$ be 
an initial data set for the evolution equations (\ref{evol1}) -- (\ref{evol4}) 
and the Vlasov equation (\ref{vlasov}) which has Bianchi symmetry and 
satisfies the constraints (\ref{cons1}) and (\ref{cons2}).
Also let $f(t_0, v)$ be a nonnegative $C^1$ function with compact support.
And assume that the potential of the scalar field $V(\phi)$ is a nonnegative $C^2$ function.
Then there exists a unique $C^1$ solution $(g_{ij}, k_{ij}, \phi, \psi, f)$ of the Einstein-Vlasov system
for all time.
\end{Theorem}
\Proof
Consider (\ref{eqn:kg_t}) with (\ref{eqn:kg^2}) 
$$
\frac{d}{dt}(k_{ij}g^{ij}) 
	= -\frac{1}{2}R + \frac{3}{2}(\sigma_{ij}\sigma^{ij}) + 
	4 \pi (T_{ij} g^{ij}) + 12\pi T_{00} + 12 \pi \psi^2
$$
Then
\begin{equation}\label{eqn:dt_kg}
\frac{d}{dt}(k_{ij}g^{ij}) \geq 0
\end{equation}
Since the cosmological models we are considering here is expanding, i.e., $(k_{ij}g^{ij}) < 0$, 
with (\ref{eqn:dt_kg}) we conclude that $(k_{ij}g^{ij})$ is bounded for $t \geq t_0$ and the proof completes
from Proposition \ref{Prop:exist2}. 
\EoPr

\subsection{Geodesic completeness with an exponential potential}\label{subsec_geo}
The next result asserts the geodesic completeness of locally spatially homogeneous spacetimes
for the Einstein-Vlasov system coupled to a nonlinear scalar field whose potential is an exponential form.
\begin{Theorem}\label{thm.geodesic}
Suppose the hypotheses of Theorem \ref{thm:exist} hold. 
And assume that the potential of the scalar field $V(\phi)$ is of form
$$V(\phi) = V_0 e^{-\lambda\kappa\phi}$$ where $V_0$ is a positive constant, $\lambda \in (0, \sqrt{2})$
and $\kappa^2 = 8\pi$.
Then the spacetime is future complete.  
\end{Theorem}
The proof of this theorem can be founded in Subsection \ref{subsec:geo}.

\section{Asymptotics of solutions with an exponential potential}\label{asymp}

We study the asymptotic behaviour of solutions in the future time 
with a particular form of the potential $V(\phi)$.
Namely the potential is given, as in the previous section, by
$V(\phi) = V_0e^{- \lambda \kappa\phi}$
where $V_0$ is a positive constant,
$0 < \lambda < \sqrt{2}$ and $\kappa^2 = 8 \pi$.
In has been shown in \cite{Hal} that in order for power-law inflation to occur 
$\lambda$ must be smaller than $\sqrt{2}$.
Note that the case $\lambda=0$ corresponds to the model
with a positive cosmological constant instead of the scalar field
which has been well understood in \cite{Lee}.
Briefly, this model exhibits exponential expansion.
For detailed information, we refer to \cite{Lee}.

We introduce a new time coordinate $\tau$ defined by
\begin{equation}\label{eqn:scaling}
d\tau = e^{-\lambda \kappa \phi /2} dt
\end{equation}
And let 
$\bar{k}_{ij} :=k_{ij}e^{\lambda\kappa\phi/2}$, 
$\bar{R} :=R e^{\lambda\kappa\phi}$,
$\bar{T}_{\alpha\beta} :=T_{\alpha\beta}e^{\lambda\kappa\phi}$ and
$\bar{\psi} := \psi e^{\lambda\kappa\phi/2}$.
Then the Hamiltonian constraint (\ref{cons1}) become
\begin{equation}
\bar{R}-(\bar{k}_{ij}\bar{k}^{ij}) + (\bar{k}_{ij}g^{ij})^2 
= 16 \pi \bar{T}_{00} + 8\pi \bar{\psi}^2 + 16 \pi V_0 \label{cons1_1}
\end{equation}
The evolution equations are
\begin{align}
\frac{d}{d\tau} g_{ij} & = -2 \bar{k}_{ij} \label{evol1_1}\\
\frac{d}{d\tau} \bar{k}_{ij} & = \bar{R}_{ij} + (\bar{k}_{lm} g^{lm}) \bar{k}_{ij}
			  			   	   - 2 \bar{k}_{il} \bar{k}^l_j - 8 \pi \bar{T}_{ij} \label{evol2_2}\\
		  			& \qquad - 4\pi \bar{T}_{00} g_{ij} + 4 \pi (\bar{T}_{lm}g^{lm}) g_{ij} 
					- 8 \pi V_0 g_{ij} +  \frac{\lambda\kappa}{2} \bar{k}_{ij}\bar{\psi} \notag\\
\frac{d}{d\tau} \phi &= \bar{\psi}, \label{evol3_3}\\
\frac{d}{d\tau} \bar{\psi} &= (\bar{k}_{lm}g^{lm}) \bar{\psi} +\lambda\kappa V_0 
			  			 	+ \frac{\lambda\kappa}{2} \bar{\psi}^2 \label{evol4_4}
\end{align}
Also the Vlasov equation becomes
\begin{equation} \label{vlasov_2}
\partial_\tau f + 
  \big\{ 2 \bar{k}^i_j v^j - e^{\lambda\kappa\phi/2}\vo^{-1/2}\gamma^i_{mn} v^m v^n \big\} \partial_{v^i}f = 0
\end{equation}

Now we define two functions :
\begin{align*}
\epsilon(\bar{\psi}, \bar{k}_{ij}g^{ij}) 
& := -\frac{2}{3}(\bar{k}_{ij}g^{ij}) - \lambda\kappa \bar{\psi}\\
\bar{S}(\bar{\psi}, \bar{k}_{ij}g^{ij})
& := (\bar{k}_{ij}g^{ij})^2 - 12 \pi(\bar{\psi}^2 + 2 V_0)
\end{align*}
Note that the function $\bar{S}$ will play the same roles as $(k_{ij}g^{ij}\pm 3\Lambda)$ 
in the papers \cite{Lee, Wald1}.

The basic idea of the following proposition is from \cite{KM2}.
Here the computation is carried out carefully so that the error terms are explicitly determined
for the future reference.

\begin{Proposition}\label{kg+sigma}
Let $\sigma_{ij}$ be the trace free part of the second fundamental form $k_{ij}$ such that
\begin{equation}  	 	 	   			 			   \label{kij}
\bar{k}_{ij} = \frac{1}{3}(\bar{k}_{lm}g^{lm}) g_{ij} + \bar{\sigma}_{ij}
\end{equation}
where $\bar{\sigma}_{ij}:= \sigma_{ij} e^{\lambda \kappa \phi/2}$
Then we have
\begin{align}
\bar{S} &= \mathcal{O}(e^{-\epsilon^*\tau}) \label{esti:barS}\\
\bar{\sigma}_{ij}\bar{\sigma}^{ij} & = \mathcal{O}(e^{-\epsilon^*\tau})\label{esti:barSigma}\\
\bar{R} &= \mathcal{O}(e^{-\epsilon^*\tau})\\
\bar{T}_{00} &=\mathcal{O}(e^{-\epsilon^*\tau})\label{esti:barT00}
\end{align}
\end{Proposition}
\Proof 
Note that using (\ref{kij}) we rewrite the constraint (\ref{cons1_1}) as
\begin{equation}\label{constrain_bar}
(\bar{k}_{ij}g^{ij})^2 = -\frac{3}{2}\bar{R} + \frac{3}{2}(\bar{\sigma}_{ij} \bar{\sigma}^{ij})
				   + 24\pi \bar{T}_{00} + 12 \pi \bar{\psi}^2 + 24 \pi V_0
\end{equation}
So $\bar{S}$ becomes
\begin{equation}\label{evol_help}
\bar{S}(\bar{\psi}, \bar{k}_{ij}g^{ij})
= -\frac{3}{2}\bar{R} + \frac{3}{2}(\bar{\sigma}_{ij} \bar{\sigma}^{ij}) + 24 \pi \bar{T}_{00}
\end{equation}  
Recall that the Ricci scalar curvature is zero or negative in all Bianchi models except type IX.
So the models are allowed to evolve in the region of $\bar{S} \geq 0$. 
Note that from (\ref{evol1_1}) and (\ref{evol2_2}) we have
$$
\frac{d}{d\tau} (\bar{k}_{ij}g^{ij})
= \bar{R} + (\bar{k}_{ij}g^{ij})^2 + 4\pi(\bar{T}_{ij}g^{ij}) -12\pi\bar{T}_{00}
-24\pi V_0 + \frac{\lambda\kappa}{2}(\bar{k}_{ij}g^{ij})\bar{\psi}
$$
Using the constraint (\ref{cons1_1}) we get
\begin{equation}\label{eqn:bar_2nd}
\frac{d}{d\tau} (\bar{k}_{ij}g^{ij})
= (\bar{k}_{ij}\bar{k}^{ij}) + 4\pi(\bar{T}_{ij}g^{ij}) +4\pi\bar{T}_{00} + 8\pi \bar{\psi}^2
-8\pi V_0 + \frac{\lambda\kappa}{2}(\bar{k}_{ij}g^{ij})\bar{\psi}
\end{equation}
Thus using (\ref{kij}) and the definitions of $\epsilon$ and $\bar{S}$, we obtain
$$
\frac{d}{d\tau} \bar{S} 
 = - \epsilon \bar{S} 
  + 2 (\bar{k}_{ij}g^{ij})[(\bar{\sigma}_{ij}\bar{\sigma}^{ij}) + 4 \pi(\bar{T}_{ij}g^{ij}) + 4\pi \bar{T}_{00}]\\
 \leq - \epsilon \bar{S}
$$
The last step is due to the fact that considering expanding spacetimes implies $(\bar{k}_{ij}g^{ij}) < 0$.
In \cite{KM2} it is shown that there exists a lower bound of $\epsilon$, say $\epsilon^*$, 
which only depends on the initial condition of the spacetimes.
Therefore we have
$$
\bar{S}(\tau) = \mathcal{O}(e^{-\epsilon^*\tau})
$$
and as a consequence from the definition of $\bar{S}$ we have
\begin{equation}\label{eqn:bar_S}
(\bar{k}_{ij}g^{ij})^2 - 12\pi (\bar{\psi}^2 +  2V_0) =\mathcal{O}(e^{-\epsilon^*\tau})
\end{equation}
By (\ref{evol_help}) the rest of the claims follows ;
$
\bar{\sigma}_{ij}\bar{\sigma}^{ij} = 
\bar{R} = 
\bar{T}_{00} = \mathcal{O}(e^{-\epsilon^*\tau}).
$
Here while we use the notation $\mathcal{O}(\cdot)$, we lose a piece of information
from above that $\bar{S}$ is non-negative.
So we want to point out that the errors in (\ref{esti:barS}) -- (\ref{esti:barT00})
are non-negative.
\EoPr

\subsection{Asymptotic behaviours of $(\bar{k}_{ij} g^{ij})$ and $\bar{\psi}$}

The estimate (\ref{eqn:bar_S}) is unsatisfactory in the sense that we do not have
sufficient information to say individual asymptotic behaviours of $(\bar{k}_{ij} g^{ij})$ and $\bar{\psi}$. 
In this subsection, we obtain asymptotic behaviours of these quantities, separately.
\begin{Proposition}\label{Prop:psi_K}
\begin{align}
\bar{\psi}& = \lambda \gamma + \mathcal{O}(e^{-\eta\tau}) \label{eqn:bar_phi}\\
\bar{k}_{ij}g^{ij} & = -3 \kappa \gamma + \mathcal{O}(e^{-\eta\tau}) \label{eqn:bar_kg}
\end{align}
where $\gamma:=\sqrt{2 V_0}/ \sqrt{6-\lambda^2}$ 
and $\eta := \min\{ \frac{1}{2} \kappa\gamma (6-\lambda^2), \epsilon^*/2\}$
\end{Proposition}
Note that in the case $0<\lambda<\sqrt{2/3}$, we have $\eta = \epsilon^*/2$.
\footnote{$\epsilon^*$, the lower bound of $\epsilon$ depends on not only the constants $\lambda$
and $V_0$ but also initial data. When $\lambda \in (0, \sqrt{2/3})$, the trivial lower bound
of $\epsilon$, which may not be sharp, is $\frac{1}{3}\kappa \sqrt{6 V_0 (2-3\lambda^2)}$ 
(see \cite{KM1,KM2} for details).
In this case $\frac{1}{2}\kappa\gamma(6-\lambda^2) > \epsilon^*/2$ is true.}

\Proof 
With the evolution equation (\ref{evol4_4}) consider
$$
\frac{d}{d\tau}(\bar{\psi}-\lambda\gamma) = F_{\bar{\psi}}(\bar{\psi}-\lambda\gamma)
$$
where 
\begin{align*}
F_{\bar{\psi}}(\bar{\psi}-\lambda\gamma) 
&:= (\bar{k}_{ij}g^{ij}+3\kappa\gamma)(\bar{\psi}-\lambda\gamma) + \frac{\lambda\kappa}{2}(\bar{\psi}-\lambda\gamma)^2\\
&\qquad + (\lambda^2-3)\kappa\gamma(\bar{\psi}-\lambda\gamma) + \lambda\gamma(\bar{k}_{ij}g^{ij}+3\kappa\gamma)
\end{align*}
(\ref{esti:barS}) implies
$$
(\bar{k}_{ij}g^{ij})^2 = \kappa^2\big[\frac{3}{2}(\bar{\psi}-\lambda\gamma)^2
					   	 +3\lambda\gamma(\bar{\psi}-\lambda\gamma)+9\gamma^2\big] +\mathcal{O}(e^{-\epsilon^*\tau})
$$
When $\bar{\psi}=\lambda\gamma$, we get
$$
\bar{k}_{ij}g^{ij} + 3 \kappa\gamma
= 3\kappa\gamma - \sqrt{9 \kappa^2\gamma^2 + \mathcal{O}(e^{-\epsilon^*\tau})} 
= \mathcal{O}(e^{-\epsilon^*\tau})
$$
Then $F_{\bar{\psi}}(0)=\mathcal{O}(e^{-\epsilon^*\tau})$ and
after a lengthy elementary computation one can see that
$F_{\bar{\psi}}'(0)= - \frac{1}{2} \kappa \gamma (6-\lambda^2) + \mathcal{O}(e^{-\epsilon^*\tau})$.
So when $\bar{\psi}$ is close to $\lambda\gamma$, 
\begin{align*}
\frac{d}{d\tau}(\bar{\psi}-\lambda\gamma)
&= \mathcal{O}(e^{-\epsilon^*\tau})
  + [-\frac{1}{2} \kappa \gamma (6-\lambda^2) + \mathcal{O}(e^{-\epsilon^*\tau})](\bar{\psi}-\lambda\gamma)\\
&\quad  + C(\bar{\psi}-\lambda\gamma)^2
\end{align*}
Define $Y_{\bar{\psi}}(\tau):=e^{\frac{1}{2} \kappa \gamma(6-\lambda^2)\tau} (\bar{\psi}-\lambda\gamma)$.
Then
$$
\frac{d}{d\tau} Y_{\bar{\psi}} 
= e^{\frac{1}{2} \kappa \gamma (6-\lambda^2)\tau}\mathcal{O}(e^{-\epsilon^*\tau})
   +\mathcal{O}(e^{-\epsilon^*\tau}) Y_{\bar{\psi}} + C e^{-\frac{1}{2} \kappa\gamma(6-\lambda^2)\tau}Y_{\bar{\psi}}^2
$$
This yields
$$
\frac{d}{d\tau} Y_{\bar{\psi}} 
= C e^{-\frac{1}{2} \kappa\gamma(6-\lambda^2)\tau}
\big(Y_{\bar{\psi}} + e^{\frac{1}{2} \kappa\gamma(6-\lambda^2)\tau}\mathcal{O}(e^{-\epsilon^*\tau/2}) \big)^2
$$
This implies that
$$
-[Y_{\bar{\psi}} + e^{\frac{1}{2} \kappa \gamma (6-\lambda^2)\tau/2} \mathcal{O}(e^{-\epsilon^*\tau/2})]^{-1} 
= C (e^{-\frac{1}{2} \kappa\gamma(6-\lambda^2)\tau} + 1)
$$
Consequently,
$$
Y_{\bar{\psi}} = C + \mathcal{O}(e^{-\frac{1}{2}\kappa\gamma(6-\lambda^2)\tau})
+ e^{\frac{1}{2}\kappa\gamma(6-\lambda^2)\tau}\mathcal{O}(e^{-\epsilon^*\tau/2})
$$
Therefore we have
$$
\bar{\psi}-\lambda\gamma = C e^{-\frac{1}{2}\kappa\gamma(6-\lambda^2)\tau} 
						   + \mathcal{O}(e^{-\kappa\gamma(6-\lambda^2)\tau}) + \mathcal{O}(e^{-\epsilon^*\tau/2})
$$
and this gives the proof of (\ref{eqn:bar_phi}).
For (\ref{eqn:bar_kg}), with (\ref{kij}) and (\ref{eqn:bar_2nd}) consider
$$
\frac{d}{d\tau}(\bar{k}_{ij}g^{ij}+3\kappa\gamma) = F_{\bar{k}}(\bar{k}_{ij}g^{ij}+3\kappa\gamma)
$$
where
\begin{align*}
&F_{\bar{k}}(\bar{k}_{ij}g^{ij}+3\kappa\gamma) \\
&\quad := \frac{1}{3}(\bar{k}_{ij}g^{ij}+3\kappa\gamma)^2 + \kappa^2 (\bar{\psi}-\lambda\gamma)^2
  	+\frac{\lambda\kappa}{2}(\bar{k}_{ij}g^{ij}+3\kappa\gamma)(\bar{\psi}-\lambda\gamma)\\
& \qquad + \frac{1}{2}\lambda\gamma\kappa^2(\bar{\psi}-\lambda\gamma) 
  		+ \frac{1}{2}(\lambda^2-4)\kappa\gamma(\bar{k}_{ij}g^{ij}+3\kappa\gamma)\\
& \qquad + (\bar{\sigma}_{ij}\bar{\sigma}^{ij}) + 4\pi \bar{T}_{00} + 4 \pi(\bar{T}_{ij}g^{ij})
\end{align*}
(\ref{esti:barS}) implies
$$
\bar{\psi}^2 = \frac{2}{3\kappa^2}(\bar{k}_{ij}g^{ij}+3\kappa\gamma)^2 
- \frac{4\gamma}{\kappa}(\bar{k}_{ij}g^{ij}+3\kappa\gamma)
+\lambda^2\gamma^2 +\mathcal{O}(e^{-\epsilon^*\tau})
$$
When $\bar{k}_{ij}g^{ij} = - 3\kappa\gamma$, one can see that
$F_{\bar{k}}(0)=\mathcal{O}(e^{-\epsilon^*\tau})$ and also
$F_{\bar{k}}'(0)=-\frac{1}{2}\kappa\gamma(6-\lambda^2)+\mathcal{O}(e^{-\epsilon^*\tau})$.
Therefore
\begin{align*}
\frac{d}{d\tau}(\bar{k}_{ij}g^{ij}+3\kappa\gamma) 
&= \mathcal{O}(e^{-\epsilon^*\tau}) 
  + [-\frac{1}{2} \kappa \lambda (6-\lambda^2) + \mathcal{O}(e^{-\epsilon^*\tau})](\bar{k}_{ij}g^{ij}+3\kappa\gamma)\\
&\qquad  + C (\bar{k}_{ij}g^{ij}+3\kappa\gamma)^2
\end{align*}
where $\bar{k}_{ij}g^{ij}+3\kappa\gamma$ is small and 
by the same argument as above (\ref{eqn:bar_kg}) follows.
\EoPr

\subsection{Relation between $\tau$ and $t$ and asymptotics in terms of $t$}\label{subsec:tau_t}
So far in the present section, we have obtain 
asymptotics of quantities, $\sigma_{ij}\sigma^{ij}$, $R$, $T_{00}$, $k_{ij}$ and $g_{ij}$,
in terms of the time coordinate $\tau$ after rescaled by a certain factor of the scalar field.
In order to study further asymptotics, it is necessary to recover these quantities
in terms of the time coordinate $t$.
For this reason,  in this subsection the relation between the two
time coordinates and the rescaling factor $e^{-\lambda\kappa\phi/2}$ in terms or $t$
will be analysed.   
\begin{Proposition} \label{prop:tau_t}
$$
e^{-\lambda^2 \kappa \gamma \tau/2}
= \frac{2 e^{\lambda \kappa C_1 /2}}{\lambda^2 \kappa \gamma}t^{-1} 
+ \left\{ 
\begin{array}{ll}
\mathcal{O}(t^{-2} \ln t), &\mbox{ if } \lambda^2 \kappa \gamma /2 = \eta \\
\mathcal{O}(t^{-(\zeta+1)}), &\mbox{ if } \lambda^2 \kappa \gamma /2 > \eta \\
\mathcal{O}(t^{-2}), &\mbox{ if } \lambda^2 \kappa \gamma /2 < \eta
\end{array}
\right.
$$
where $\zeta := 2\eta/ \lambda^2 \kappa \gamma$ and $C_1$ is a constant.
\end{Proposition}
\Proof 
From (\ref{eqn:bar_phi}) we have
\begin{equation}\label{eqn:phi_tau}
\phi(\tau) = \lambda \gamma \tau + C_1 + \mathcal{O}(e^{-\eta\tau})
\end{equation}
So by (\ref{eqn:scaling}) we get
$$
t(\tau) = t(\tau_0) 
+  e^{\lambda\kappa C_1/2} \int^\tau_{\tau_0} 
\exp[\lambda^2 \kappa \gamma s /2+ \mathcal{O}(e^{-\eta s})] \,ds
$$
where $t(\tau_0) = t_0$.
Then
\begin{align*}
& e^{-\lambda^2 \kappa \gamma \tau/2} t(\tau)\\ 
&\quad = e^{-\lambda^2 \kappa \gamma \tau/2} t(\tau_0) 
 + e^{\lambda\kappa C_1/2} e^{-\lambda^2 \kappa \gamma \tau /2} 
  \int^\tau_{\tau_0} \exp[\lambda^2 \kappa \gamma s /2 + \mathcal{O}(e^{-\eta s})] \,ds\\
& \quad =e^{-\lambda^2 \kappa \gamma \tau/2} t(\tau_0) 
  + e^{\lambda\kappa C_1/2} \int^\tau_{\tau_0} e^{\lambda^2 \kappa \gamma (s-\tau)/2}[1 + \mathcal{O}(e^{-\eta s})] \,ds\\
& \quad = \frac{2 e^{\lambda \kappa C_1 /2}}{\lambda^2 \kappa \gamma}
+
\left\{ 
\begin{array}{ll}
\mathcal{O}(\tau e^{-\lambda^2 \kappa \gamma \tau/2}),  &\mbox{ if } \lambda^2 \kappa \gamma /2 = \eta \\
\mathcal{O}(e^{-\eta\tau}),
   	 			 &\mbox{ if } \lambda^2 \kappa \gamma /2 > \eta\\
\mathcal{O}(e^{-\lambda^2 \kappa \gamma \tau/2}),
   	 			 &\mbox{ if } \lambda^2 \kappa \gamma /2 < \eta
\end{array}
\right.
\end{align*}
In all cases, we have
$$
e^{-\lambda^2 \kappa \gamma \tau /2} t(\tau) \leq  C
$$
Consequently
$$
e^{-\lambda^2 \kappa \gamma \tau /2} = \mathcal{O}(t^{-1})
$$
Thus the proposition follows.
\EoPr

\begin{Proposition}\label{prop:e_phi}
$$
e^{-\lambda \kappa \phi /2}
= 
\frac{2}{\lambda^2 \kappa \gamma}t^{-1} 
+ \left\{ 
\begin{array}{ll}
\mathcal{O}(t^{-2} \ln t), &\mbox{ if } \lambda^2 \kappa \gamma /2 = \eta \\
\mathcal{O}(t^{-(\zeta+1)}), &\mbox{ if } \lambda^2 \kappa \gamma /2 > \eta \\
\mathcal{O}(t^{-2}), &\mbox{ if } \lambda^2 \kappa \gamma /2 < \eta
\end{array}
\right.
$$
\end{Proposition}
\Proof
By (\ref{eqn:phi_tau}),
$$
e^{-\lambda\kappa\phi/2} = e^{-\lambda^2\kappa\gamma\tau/2}e^{-\lambda\kappa C_1/2}(1+\mathcal{O}(e^{-\eta\tau}))
$$
Combining this with Proposition \ref{prop:tau_t} yields the conclusion of the proposition.
\EoPr

\begin{Proposition}
\begin{align}
\sigma_{ij}\sigma^{ij} & = \mathcal{O}(t^{-(\xi+2)}) \label{esti:sigma_t}\\
R &= \mathcal{O}(t^{-(\xi+2)}) \label{esti:R_t}\\
T_{00} &=\mathcal{O}(t^{-(\xi+2)}) \label{esti_T_t}
\end{align}
where $\xi := 2\epsilon^*/\lambda^2\kappa\gamma$.
\end{Proposition}
\Proof 
By Proposition \ref{prop:tau_t}
$$
e^{-\epsilon^*\tau} = \mathcal{O}(t^{-2\epsilon^*/\lambda^2\kappa\gamma\tau})
$$
Also Proposition  \ref{prop:e_phi} implies
$$
e^{-\lambda\kappa\phi} = \mathcal{O}(t^{-2})
$$ 
Combining these with (\ref{esti:barSigma}) -- (\ref{esti:barT00}) in 
Proposition \ref{kg+sigma} concludes the proposition.
\EoPr

\subsection{Asymptotic behaviours of $k_{ij}g^{ij}$, $\psi$ and $\phi$ in terms of $t$}
In this part, we will observe asymptotic behaviours of $k_{ij}g^{ij}$, $\psi$ and $\phi$ in terms of $t$
using the relation between two time coordinates $\tau$ and $t$ we have obtained in the previous subsection.
\begin{Proposition}\label{prop:phi_kg}
\begin{align} 	   
\psi& = \frac{2}{\lambda\kappa}t^{-1}\label{prop:phi} 
+ \left\{ 
\begin{array}{ll}
\mathcal{O}(t^{-2} \ln t), &\mbox{ if } \lambda^2 \kappa \gamma /2 = \eta \\
\mathcal{O}(t^{-(\zeta+1)}), &\mbox{ if } \lambda^2 \kappa \gamma /2 > \eta \\
\mathcal{O}(t^{-2}), &\mbox{ if } \lambda^2 \kappa \gamma /2 < \eta
\end{array}
\right.\\
k_{ij}g^{ij} & = -\frac{6}{\lambda^2}t^{-1}\label{prop:kg}
+ \left\{ 
\begin{array}{ll}
\mathcal{O}(t^{-2} \ln t), &\mbox{ if } \lambda^2 \kappa \gamma /2 = \eta \\
\mathcal{O}(t^{-(\zeta+1)}), &\mbox{ if } \lambda^2 \kappa \gamma /2 > \eta \\
\mathcal{O}(t^{-2}), &\mbox{ if } \lambda^2 \kappa \gamma /2 < \eta
\end{array}
\right.
\end{align}
\end{Proposition}
\Proof 
Note that from Proposition \ref{prop:tau_t} 
$$e^{-\eta \tau} = \mathcal{O}(t^{-\zeta})$$
So combining (\ref{eqn:bar_phi}) in Proposition \ref{Prop:psi_K} 
and Proposition \ref{prop:e_phi}, (\ref{prop:phi}) follows.
With (\ref{eqn:bar_kg}) in Proposition \ref{Prop:psi_K} 
the same argument applies to prove (\ref{prop:kg}).
\EoPr

\begin{Proposition}
$$
\phi = \frac{2}{\lambda\kappa}\ln t 
+ C_2 + \left\{ 
\begin{array}{ll}
\mathcal{O}(t^{-1} \ln t), &\mbox{ if } \lambda^2 \kappa \gamma /2 = \eta \\
\mathcal{O}(t^{-\zeta}), &\mbox{ if } \lambda^2 \kappa \gamma /2 > \eta \\
\mathcal{O}(t^{-1}), &\mbox{ if } \lambda^2 \kappa \gamma /2 < \eta
\end{array}
\right.
$$
where $C_2$ is a constant.
\end{Proposition}
\Proof
The proposition follows directly from (\ref{prop:phi}) in the previous proposition.
\EoPr

\subsection{Asymptotic behaviours of $g_{ij}$ and $g^{ij}$}
Asymptotics of $g_{ij}$ and $g^{ij}$ will be analysed first in the time coordinate $\tau$.
By means of the relation between the two time coordinates $\tau$ and $t$ 
in Subsection \ref{subsec:tau_t}, these asymptotics will be recovered in terms of $t$.

It has been observed in Proposition \ref{kg+sigma} 
that $\bar{\sigma}_{ij}\bar{\sigma}^{ij} = \mathcal{O}(e^{-\epsilon^*\tau})$.
Using the following two lemmas we will identify $\bar{\sigma}_{ij}(\tau)$
which play a role to analyse $g_{ij}(\tau)$.
\begin{Lemma}\label{lem:sigma}
Let $\|g(\tau)\|$, $\|\bar{k}(\tau)\|$ and $\|\bar{\sigma}(\tau)\|$ 
denote the norms of the matrices with entries 
$g_{ij}(\tau)$, $\bar{k}_{ij}(\tau)$ and $\bar{\sigma}_{ij}(\tau)$, respectively. Then
\begin{equation*}
\|\bar{\sigma}(\tau)\| \leq C e^{-\epsilon^*\tau/2} \| g(\tau) \|
\end{equation*}
\end{Lemma}
\Proof
The lemma follows by the fact that
\begin{equation*}
\|\bar{\sigma}(\tau)\|_{g(\tau)} \leq (\bar{\sigma}_{ij} \bar{\sigma}^{ij})^{1/2} \leq C e^{-\epsilon^*\tau/2}
\end{equation*}
The last inequality is due to (\ref{esti:barSigma}).
\EoPr

\begin{Lemma} \label{egij}
\begin{align}
|e^{-2\kappa\gamma\tau} g_{ij}(\tau)| & \leq C\\
|e^{2\kappa\gamma\tau} g^{ij}(\tau)| & \leq C \label{2ndgij}
\end{align}
\end{Lemma}
\Proof
Let $\tilde{g}_{ij}(\tau) := e^{-2\kappa\gamma\tau} g_{ij}(\tau)$. 
Then using (\ref{kij}) and (\ref{eqn:bar_kg}),
we get
\begin{align}
\frac{d}{d\tau} \tilde{g}_{ij} 
&= -2\kappa\gamma \tilde{g}_{ij} - \frac{2}{3}(\bar{k}_{lm} g^{lm}) \tilde{g}_{ij} 
   - 2 e^{-2 \kappa\gamma\tau} \bar{\sigma}_{ij} \notag\\
&= \mathcal{O}(e^{-\eta\tau}) \tilde{g}_{ij} - 2 e^{-2\kappa\gamma\tau}\bar{\sigma}_{ij} \label{bargij}		   				
\end{align}
Now let us use the norms again. Let $\|\tilde{g}(\tau)\|$ be a norm of the matrix with entries 
$\tilde{g}_{ij}(\tau)$. Then we get
\begin{align}\label{eqn_gtau}
\|\tilde{g}(\tau)\| &\leq \|\tilde{g}(\tau_0)\| + \int^\tau_{\tau_0}  \big(C e^{-\eta s} \|\tilde{g}(s)\|
				 	    + C e^{-2\kappa\gamma s}\|\bar{\sigma}(s)\| \big)\, ds\\
			   &\leq \|\tilde{g}(\tau_0)\| + \int^\tau_{\tau_0}  \big(C e^{-\eta s} \|\tilde{g}(s)\| 
			   		 	+ C e^{-2\kappa\gamma s} \|\bar{\sigma}(s)\|_{\tilde{g}}\|\tilde{g}(s)\|\big)\, ds\notag
\end{align}
Note that
\begin{equation*}
\|\bar{\sigma}(\tau)\|_{\tilde{g}} \leq e^{2\kappa\gamma\tau}(\bar{\sigma}_{ij} \bar{\sigma}^{ij})^{1/2}
\leq C e^{2\kappa\gamma\tau}e^{-\epsilon^*\tau/2}
\end{equation*}
So combining this with (\ref{eqn_gtau}) yields
\begin{equation*}
\|\tilde{g}(\tau)\| \leq \|\tilde{g}(\tau_0)\| + \int^\tau_{\tau_0}  C e^{- \eta s} \|\tilde{g}(s)\| \, ds
\end{equation*}
By Gr\"onwall's inequality, this becomes
\begin{equation*}
\|\tilde{g}(\tau)\| \leq \|\tilde{g}(\tau_0)\| 
					\exp\left[ \int^\tau_{\tau_0} C e^{- \eta s} \, ds \right] \leq C
\end{equation*}
Therefore $|\tilde{g}_{ij}(\tau)|$ is bounded by a constant for all $\tau \geq \tau_0$. 
Also (\ref{2ndgij}) follows by the same argument.
\EoPr

\begin{Proposition}\label{sigmaijt}
\begin{equation*}
e^{-2\kappa\gamma\tau}\bar{\sigma}_{ij}= \mathcal{O} (e^{-\epsilon^*\tau/2})
\end{equation*}
\end{Proposition}
\Proof
The proposition follows by Lemmas \ref{lem:sigma} and \ref{egij}.
\EoPr

\begin{Proposition}\label{prop_g}
\begin{align}
g_{ij}(\tau) &= e^{2\kappa\gamma\tau}(\mathcal{G}_{ij} + \mathcal{O}(e^{-\eta\tau})) \label{g_ij}\\
g^{ij}(\tau) &= e^{-2\kappa\gamma\tau}(\mathcal{G}^{ij} + \mathcal{O}(e^{-\eta\tau})) \label{g^ij}
\end{align}
Here $\mathcal{G}_{ij}$ and $\mathcal{G}^{ij}$ are independent of $\tau$.
\end{Proposition}
\Proof
Let us consider (\ref{bargij}) again ;
$$
\frac{d}{d\tau} \tilde{g}_{ij}(\tau) 
	   = \mathcal{O}(e^{-\eta\tau}) \tilde{g}_{ij}(\tau) - 2 e^{-2 \kappa\gamma\tau} \bar{\sigma}_{ij}(\tau)
$$
where $\tilde{g}_{ij}(\tau) = e^{-2\kappa\gamma\tau} g_{ij}(\tau)$.
Then Lemma \ref{egij} and Proposition \ref{sigmaijt} imply
$$
\frac{d}{d\tau} \tilde{g}_{ij}(\tau) 
	  \leq C e^{- \eta\tau}\tilde{g}_{ij} + C e^{-\epsilon^*\tau/2}
	  \leq C e^{- \eta\tau}
$$
Since $\frac{d}{d\tau} \tilde{g}_{ij}$ is decaying exponentially,
there exists a limit, say $\mathcal{G}_{ij}$ , of $\tilde{g}_{ij}$ as $\tau$ goes to infinity.
Then this gives
\begin{equation*}
\tilde{g}_{ij}(\tau) = \mathcal{G}_{ij} + \mathcal{O}(e^{- \eta\tau})
\end{equation*}
i.e.
\begin{equation*}
g_{ij}(\tau) = e^{2\kappa\gamma\tau} \big( \mathcal{G}_{ij} + \mathcal{O}(e^{- \eta\tau}) \big)
\end{equation*}
Here the lower order term of $g_{ij}(\tau)$ is of an exponential form
so that it is combined with the leading order term, which makes it possible to compute $g^{ij}(\tau)$ explicitly. 
So $g^{ij}(\tau)$ is
$$
g^{ij}(\tau) = e^{-2\kappa\gamma\tau}\big(\mathcal{G}^{ij}+\mathcal{O}(e^{-\eta\tau})\big)
$$
\EoPr

\begin{Proposition}\label{prop_gt}
\begin{align}
g_{ij}(t) &= t^{4/\lambda^2}\left( 
\Big(\frac{\lambda^2\kappa\gamma}{2e^{\lambda\kappa C_1/2}}\Big)^{4/\lambda^2} \mathcal{G}_{ij}
+ 
\left\{ 
\begin{array}{ll}
\mathcal{O}(t^{-1} \ln t), &\mbox{ if } \lambda^2 \kappa \gamma /2 = \eta \\
\mathcal{O}(t^{-\zeta}), &\mbox{ if } \lambda^2 \kappa \gamma /2 > \eta \\
\mathcal{O}(t^{-1}), &\mbox{ if } \lambda^2 \kappa \gamma /2 < \eta
\end{array}
\right. \right)  \label{g_ijt}\\
g^{ij}(t) &= t^{-4/\lambda^2}\left( 
\Big(\frac{2e^{\lambda\kappa C_1/2}}{\lambda^2\kappa\gamma}\Big)^{4/\lambda^2} \mathcal{G}^{ij}
+ 
\left\{ 
\begin{array}{ll}
\mathcal{O}(t^{-1} \ln t), &\mbox{ if } \lambda^2 \kappa \gamma /2 = \eta \\
\mathcal{O}(t^{-\zeta}), &\mbox{ if } \lambda^2 \kappa \gamma /2 > \eta \\
\mathcal{O}(t^{-1}), &\mbox{ if } \lambda^2 \kappa \gamma /2 < \eta
\end{array}
\right. \right)
\label{g^ijt}
\end{align}
\end{Proposition}
\Proof
Recall that 
$$
e^{-\eta\tau} = \mathcal{O}(t^{-\zeta})
$$
Proposition \ref{prop:tau_t} implies
\begin{align*}
e^{2 \kappa\gamma\tau} &= t^{4/\lambda^2} \left(
\frac{2e^{\lambda\kappa C_1/2}}{\lambda^2\kappa\gamma}
+\left\{ 
\begin{array}{ll}
\mathcal{O}(t^{-1} \ln t), &\mbox{ if } \lambda^2 \kappa \gamma /2 = \eta \\
\mathcal{O}(t^{-\zeta}), &\mbox{ if } \lambda^2 \kappa \gamma /2 > \eta \\
\mathcal{O}(t^{-1}), &\mbox{ if } \lambda^2 \kappa \gamma /2 < \eta
\end{array}
\right.
\right)^{-4/\lambda^2}\\
&= t^{4/\lambda^2} \left(
\Big(\frac{\lambda^2\kappa\gamma}{2e^{\lambda\kappa C_1/2}}\Big)^{4/\lambda^2}
+\left\{ 
\begin{array}{ll}
\mathcal{O}(t^{-1} \ln t), &\mbox{ if } \lambda^2 \kappa \gamma /2 = \eta \\
\mathcal{O}(t^{-\zeta}), &\mbox{ if } \lambda^2 \kappa \gamma /2 > \eta \\
\mathcal{O}(t^{-1}), &\mbox{ if } \lambda^2 \kappa \gamma /2 < \eta
\end{array}
\right.
\right)
\end{align*}
So by means of this and Proposition \ref{prop_g}, (\ref{g_ijt}) follows.
The same argument for $ g^{ij}$ yields (\ref{g^ijt}).
\EoPr

\subsection{Asymptotics of the generalized Kasner exponents and the deceleration parameter}
Let $\lambda_i$ be the eigenvalues of $k_{ij}(t)$ with respect to $g_{ij}(t)$, 
i.e., the solutions of $\det(k^i_j - \lambda \delta^i_j)=0$.
Define {\em the generalized Kasner exponents} by
$$p_i := \frac{\lambda_i}{\sum_l \lambda_l} = \frac{\lambda_i}{(k_{lm}g^{lm})}$$
The name comes from the fact that in the special case of the Kasner solutions
these are the Kasner exponents. Note that while the Kasner exponents are constants,
the generalized Kasner exponents are in general functions of $t$.
The generalized Kasner exponents always satisfy the first of the two Kasner
relations, but in general do not satisfy the second,
where these two Kasner relations are
\begin{gather}
\sum_i p_i = 1, \label{kasner_rel1}\\
\sum_i (p_i)^2 = 1.
\end{gather}
The following proposition exhibits that the spacetime isotropizes at late times.

\begin{Proposition}
$$
p_i(t) = \frac{1}{3} + \mathcal{O}(t^{-\xi/2})
$$
where $\xi = 2\epsilon^*/\lambda^2\kappa\gamma$.
\end{Proposition}
\Proof
First note that by (\ref{kij}) $\lambda_i$ are also the solutions of
$$
\det\big( \bar{\sigma}^i_j - [\lambda - \frac{1}{3} (\bar{k}_{lm}g^{lm})] \delta^i_j\big)=0
$$
So the eigenvalues of $\bar{\sigma}_{ij}(\tau)$ with respect to $g_{ij}(\tau)$ are
$$\tilde{\lambda}_i := \lambda_i - \frac{1}{3} (\bar{k}_{lm}g^{lm})$$
Also note that $\sum_i (\tilde{\lambda}_i)^2 = \bar{\sigma}_{lm} \bar{\sigma}^{lm}$.
Then (\ref{esti:barSigma}) implies
$$\tilde{\lambda}_i = \mathcal{O}(e^{-\epsilon^*\tau/2})$$
Therefore using this and (\ref{eqn:bar_kg}) we obtain
$$p_i - \frac{1}{3} = \frac{\tilde{\lambda}_i}{\frac{1}{3}(\bar{k}_{lm}g^{lm})} = \mathcal{O}(e^{-\epsilon^*\tau/2})$$
Thus Proposition \ref{prop:tau_t} completes the proof.
\EoPr

There is another quantity to be considered regarding expanding cosmological models, 
which is {\em the deceleration parameter}, say $q$. This deceleration
parameter is related to the mean curvature, as follows
$$
\frac{d}{dt} (k_{ij} g^{ij}) = - (1+ q)(k_{ij} g^{ij})^2
$$
In accelerated expanding models, the deceleration parameter is negative.

\begin{Proposition}
$$ q = -1 - \frac{\lambda^2}{6}
+ \left\{ 
\begin{array}{ll}
\mathcal{O}(t^{-1} \ln t), &\mbox{ if } \lambda^2 \kappa \gamma /2 = \eta \\
\mathcal{O}(t^{-\zeta}), &\mbox{ if } \lambda^2 \kappa \gamma /2 > \eta \\
\mathcal{O}(t^{-1}), &\mbox{ if } \lambda^2 \kappa \gamma /2 < \eta
\end{array}
\right.
$$
\end{Proposition}
\Proof
The proof is a straight forward computation from (\ref{prop:kg}).
\EoPr

\subsection{Asymptotics of momenta}\label{subsec:v}
We will analyse the behaviour of the momenta of the distribution function $f$
along the characteristics where $f$ is a constant.
From the Vlasov equation (\ref{vlasov})
we define the characteristic curve $V^i (t)$ by
\begin{equation}
\frac{dV^i}{dt} = 2 k^i_j V^j - (1+g_{rs}V^r V^s)^{-1/2}\gamma^i_{mn} V^m V^n
\end{equation}
for each $V^i(t_0) = v^i_0$ given $t_0$.
The characteristics $V_i$, rather than $V^i$, has a simpler form,
which makes analysing the behaviour of the momenta easier.
So here $V_i (t)$ satisfies
\begin{equation}\label{dVi}
\frac{dV_i}{dt} = - (1+g_{rs}V^r V^s)^{-1/2}\gamma^j_{mn} V_{p} V_{q}
				  g^{pm}g^{qn}g_{ij}
\end{equation}
for each $V_i(t_0) = v_{i0}$ given $t_0$. 
Also observe that $V_i(\tau)$ satisfies
\begin{equation}\label{dVi_tau}
\frac{dV_i}{d\tau} = - e^{\lambda\kappa\phi/2}\vo^{-1/2}\gamma^i_{mn}  V_{p} V_{q}
				  g^{pm}g^{qn}g_{ij}
\end{equation}
For the rest of the paper, the capital $V^i$ and $V_i$ indicate that $v^i$ and $v_i$
are parameterized by the coordinate time $t$ or $\tau$.

\begin{Theorem} \label{thm:V}
$V_i(t)$ from (\ref{dVi}) converges to a constant along the characteristics 
as $t$ goes to infinity. That is
$$
V_i(t) = C_3 +
\left\{ 
\begin{array}{ll}
\mathcal{O}(t^{-\zeta}), &\mbox{ if } \zeta \leq 1, \\
\mathcal{O}(t^{-\omega}), &\mbox{ if } \zeta >1,
\end{array}
\right. 
$$
where $\omega := \min \{\zeta, 4/\lambda^2 -1\}$.
Furthermore, 
$$V^i(t) = t^{-4/\lambda^2} \left(C_4 + 
\left\{ 
\begin{array}{ll}
\mathcal{O}(t^{-1} \ln t), &\mbox{ if } \lambda^2 \kappa \gamma /2 = \eta \\
\mathcal{O}(t^{-\zeta}), &\mbox{ if } \lambda^2 \kappa \gamma /2 > \eta \\
\mathcal{O}(t^{-1}), &\mbox{ if } \lambda^2 \kappa \gamma /2 < \eta
\end{array}
\right. 
\right)$$ 
\end{Theorem}

Before the proof of this theorem, some lemmas are required.

\begin{Lemma}\label{gvv_tau}
$$
(g^{ij}V_i V_j)(\tau) = e^{-2 \kappa\gamma\tau} \big(\mathcal{V} + \mathcal{O}(e^{- \eta\tau})\big)
$$
where $\mathcal{V}$ is a constant.
\end{Lemma}
\Proof
First note that by Propositions \ref{sigmaijt} and \ref{prop_g}, we have
$$\bar{\sigma}^{ij} = \mathcal{O}(e^{-(2\kappa\gamma + \epsilon^*/2)\tau})$$
Let $\tilde{\sigma}^{ij}:= e^{(2\kappa\gamma + \epsilon^*/2)\tau} \bar{\sigma}^{ij}$.
Then $\tilde{\sigma}^{ij}$ is bounded by a constant for all $\tau \geq \tau_0$.
Since $\mathcal{G}^{ij}$ in Proposition \ref{prop_g} is positive definite and time independent,
there exists a constant $C$, independent of time, such that
$$
\tilde{\sigma}^{ij} V_i V_j \leq C \mathcal{G}^{ij} V_i V_j
$$
Then by means of this and (\ref{eqn:bar_kg}), we obtain
\begin{align}
\frac{d}{d\tau} (g^{ij} V_i V_j )
					 &= 2 \bar{k}^{ij} V_i V_j \notag\\
				  	 &= \frac{2}{3} (\bar{k}_{lm}g^{lm}) g^{ij} V_i V_j 
						   	  + 2 \bar{\sigma}^{ij} V_i V_j\notag\\
					 & \leq \big( -2 \kappa \gamma + Ce^{-\eta\tau} \big) g^{ij} V_i V_j 
					   		+ C e^{-(2\kappa\gamma + \epsilon^*/2)\tau} \mathcal{G}^{ij} V_i V_j \notag\\
					 & \leq \big(-2\kappa\gamma + Ce^{- \eta\tau} \big)g^{ij} V_i V_j \label{dsgvv}
\end{align}
Here to get the first equal sign, (\ref{dVi_tau}) is used. Yet the terms involved
with (\ref{dVi_tau}) vanish due to the antisymmetric property of $\gamma^l_{mn}$ combining with $g^{ij}$.
Now consider $V_\tau := e^{2 \kappa\gamma\tau} g^{ij} V_i V_j$.
Then one can see from (\ref{dsgvv}) that
$$
\frac{dV_\tau}{d\tau} = \mathcal{O}(e^{- \eta\tau}) V_\tau
$$
So there exists a limit of $V_\tau$, say $\mathcal{V}$ as $\tau$ goes to infinity.
Then we have
$$
V_\tau = \big(\mathcal{V}+\mathcal{O}(e^{-\eta\tau})\big)
$$
This completes the proof.
\EoPr

\begin{Lemma}\label{gvv}
$$
(g^{ij}V_i V_j)(t)
= \mathcal{V}\left(\frac{2e^{\lambda\kappa C_1/2}}{\lambda^2\kappa\gamma} \right)^{4/\lambda^2} t^{-4/\lambda^2}
+ \left\{ 
\begin{array}{ll}
\mathcal{O}(t^{-2} \ln t), &\mbox{ if } \lambda^2 \kappa \gamma /2 = \eta \\
\mathcal{O}(t^{-(\zeta+1)}), &\mbox{ if } \lambda^2 \kappa \gamma /2 > \eta \\
\mathcal{O}(t^{-2}), &\mbox{ if } \lambda^2 \kappa \gamma /2 < \eta
\end{array}
\right.
$$
\end{Lemma}
\Proof
Recall that
$$
e^{-\eta\tau} = \mathcal{O}(t^{-\zeta})
$$
Then by means of Proposition \ref{prop:tau_t} and Lemma \ref{gvv_tau} we have
\begin{align*}
(g^{ij}V_i V_j)(t)& = \mathcal{V}\left(\frac{2e^{\lambda\kappa C_1/2}}{\lambda^2\kappa\gamma} t^{-1}\right)^{4/\lambda^2}
		   		   + \mathcal{O}(t^{-(\zeta + \frac{4}{\lambda^2})}) \\
				  & \qquad + \left\{ 
\begin{array}{ll}
\mathcal{O}(t^{-2} \ln t), &\mbox{ if } \lambda^2 \kappa \gamma /2 = \eta \\
\mathcal{O}(t^{-(\zeta+1)}), &\mbox{ if } \lambda^2 \kappa \gamma /2 > \eta \\
\mathcal{O}(t^{-2}), &\mbox{ if } \lambda^2 \kappa \gamma /2 < \eta
\end{array}
\right.
\end{align*}
Note that $\zeta + 4/\lambda^2 > \zeta + 2$. So the lemma follows.
\EoPr

\begin{Lemma}\label{vi}
$$
|V_i|(t)\leq C_5 + \mathcal{O}(t^{-\zeta})
$$
where $C_5$ is a constant and for all $i$.
\end{Lemma}
\Proof
Since $\mathcal{G}^{ij}$ is positive definite, there exists a constant
$C$ such that 
$$
|V_i|^2(\tau) \leq C \mathcal{G}^{ij} V_i(\tau) V_j(\tau)
$$ 
Note that $e^{-2\kappa\gamma\tau} \mathcal{G}^{ij} V_i V_j$ is the leading order term 
in $(g^{ij} V_i V_j)(\tau)$ in (\ref{g^ij}). So using Lemma \ref{gvv_tau}, we conclude that
$$
\mathcal{G}^{ij} V_i V_j = \mathcal{V} + \mathcal{O}(e^{-\eta\tau})
$$
Combining this with the fact that
$$
e^{-\eta\tau} = \mathcal{O}(t^{-\zeta})
$$
we complete the proof. 
\EoPr

{\sc {Proof of Theorem \ref{thm:V}}}. 
Note that in Bianchi type I, since all structure constants are zero,  
also the Ricci rotation coefficients $\gamma^j_{mm}$ are zero. 
So from (\ref{dVi}) it is clear that $V_i (t) = v_{i0}$ for all $t$.

More generally Lemma \ref{vi} says that all $V_i$ are bounded
by a constant when $t$ goes to infinity. However this allows 
oscillating behaviours. So to rule out these cases, it is necessary to analyse $\frac{dV_i}{dt}$.
Combining Proposition \ref{prop_gt} and (\ref{dVi}), we have
\begin{equation*}
\left|\frac{dV_i}{dt} \right| \leq C t^{-4/\lambda^2} |V_p| |V_q|
\end{equation*}
Here the right hand side is a summation for some $p$ and $q$.
Then by Lemma \ref{vi} this implies
\begin{equation}\label{eqn:dv_dt}
\left|\frac{dV_i}{dt}\right| \leq C t^{-4/\lambda^2}
\end{equation}
This leads to the conclusion that when $t$ goes to the infinity, 
$\frac{dV_i}{dt}$ goes to zero, and so $V_i$ goes to a constant.
Combining Lemma \ref{vi} and (\ref{eqn:dv_dt}) we have
$$
V_i (t) = C_3 + \mathcal{O}(t^{-\zeta}) + \mathcal{O}(t^{-(4/\lambda^2 -1)})
$$ 
Since $4/\lambda^2-1 >1$
$$
V_i (t) = C_3 +
\left\{ 
\begin{array}{ll}
\mathcal{O}(t^{-\zeta}), &\mbox{ if } \zeta \leq 1, \\
\mathcal{O}(t^{-\omega}), &\mbox{ if } \zeta > 1,
\end{array}
\right. 
$$
where $\omega = : \min\{4/\lambda^2-1, \zeta\}$.
Now combining this with (\ref{g^ijt}) completes the proof.
\EoPr

\subsection{Geodesic completeness}\label{subsec:geo}
In this part, we will prove the completeness of future directed causal geodesics,
which has been postponed in Subsection \ref{subsec_geo}.

{\sc Proof of Theorem \ref{thm.geodesic}.}
The geodesic equations for a metric of the form (\ref{metric}) imply that
along the geodesics the variables $t$, $v^i$ and $v^0$ satisfy the following
system of differential equations :
\begin{align}
\frac{dt}{ds} &= v^0 \label{geoeq1}\\
\frac{dv^0}{ds} & = k_{ij} v^i v^j \notag\\
\frac{dv^i}{ds} &= 2 k^i_j v^j v^0 - \gamma^i_{mn} v^m v^n \notag
\end{align}
where $s$ is an affine parameter. For a particle with rest mass $m$ moving forward in time,
$v^0$ can be expressed by the remaining variables,
\begin{equation}\label{geoeq2}
 v^0 = (m^2 + g_{ij}v^i v^j)^{1/2}
\end{equation}

The geodesic completeness is decided by looking at the relation between 
$t$ and the affine parameter $s$,
along any future directed causal geodesic. This relation is clear from (\ref{geoeq1}) and
(\ref{geoeq2}). I.e., it is given by
\begin{equation*} 
\frac{dt}{ds} = (m^2 + g_{ij} v^i v^j)^{1/2}
\end{equation*}
To control this, it is necessary to control $g_{ij} v^i v^j$ as a function of the coordinate time $t$.
Consider first the case of a timelike geodesic. I.e., $m >0$.
Then $V^i(t)$ satisfy
$$
\frac{dV^i}{dt} = 2 k^i_j V^j- (m^2 + g_{ij}V^i V^j)^{-1/2} \gamma^i_{mn} V^m V^n
$$
In this case, the arguments presented in Subsection \ref{subsec:v} is valid, 
in particular those in Lemmas \ref{gvv_tau} and \ref{gvv} when $m=1$.
Therefore by Lemma \ref{gvv}, $(g_{ij} V^i V^j)(t)$ is bounded above by $C t^{-4/\lambda^2}$,
and so by $C$, for all $t \geq t_0$.
Hence this gives that 
$$
\frac{ds}{dt} = \big(m^2 + (g_{ij} V^i V^j)(t)\big)^{-1/2} \geq  C
$$
Therefore when $s$ is recovered by integrating this, 
the integral of the right hand side diverges as $t$ goes to infinity.

Now consider a null geodesic, i.e., $m=0$.
Then in this case $V^i(t)$ satisfy
$$
\frac{dV^i}{dt} = 2 k^i_j V^j - (g_{ij}V^i V^j)^{-1/2} \gamma^i_{mn} V^m V^n
$$
Also Lemma \ref{gvv} is valid.
Therefore $(g_{ij} V^i V^j)(t)$ is bounded by a constant and this gives
\begin{equation*}
\frac{ds}{dt} = \big(g_{ij} (V^i V^j)(t)\big)^{-1/2} \geq  C
\end{equation*}
Therefore as $t$ goes to infinity so does $s$.
\EoPr

\subsection{Asymptotics of the energy-momentum tensor}
In this subsection it will be analysed the asymptotic behaviour of the energy-momentum tensor
in an orthonormal frame on the hypersurfaces.
\begin{Proposition}
Let $\{\hat{e}_i\}$ be an orthonormal frame. 
The energy-momentum tensor is described by
\begin{align}
\rho (t) & = \int f(t, \hat{v}) (1+|\hat{v}|^2)^{1/2} \,d\hat{v}\\
J_i (t) & = \int f(t, \hat{v}) \hat{v}_i \,d\hat{v}\\
S_{ij} (t) & = \int f(t, \hat{v}) \hat{v}_i \hat{v}_j (1+|\hat{v}|^2)^{-1/2} \,d\hat{v}
\end{align}
where $\rho:= \hat{T}_{00}$ is the energy density, $J_i := \hat{T}_{0i}$ the components of the current density
and $S_{ij} := \hat{T}_{ij}$ are the spatial components of the energy-momentum tensor.
Here the hats indicate that objects are written in the orthonormal frame.
Furthermore $\hat{v} := (\hat{v}_1, \hat{v}_2, \hat{v}_3)$ and $d \hat{v} = d\hat{v}_1 d\hat{v}_2 d\hat{v}_3$.

Then $\rho (t)$, $J_i(t)$ and $S_{ij}(t)$ tend to zero as $t$ goes to infinity.
More precisely, 
\begin{align*}
\rho (t) &= \mathcal{O}(t^{-6/\lambda^2})\\
J_i(t) &= \mathcal{O}(t^{-8/\lambda^2})\\
S_{ij}(t) &= \mathcal{O}(t^{-10/\lambda^2})
\end{align*}
Furthermore
\begin{align}
\frac{J_i(t)}{\rho(t)} &= \mathcal{O}(t^{-2/\lambda^2})\label{ratio1}\\
\frac{S_{ij}(t)}{\rho(t)} &= \mathcal{O}(t^{-4/\lambda^2}) \label{ratio2}
\end{align}
\end{Proposition}
\Proof
Note that $f(t_0, v)$ has compact support on $v$.
Also observe that Theorem \ref{thm:V} implies that $V_i(t)$ is {\em uniformly bounded}.
Combining these two facts implies that there exists a constant $C$ such that 
\begin{equation}\label{f_vi}
f(t, v)=0, \quad \text{if } |v_i| \geq C 
\end{equation}
for all $t$.
By (\ref{g_ijt}) we have
\begin{equation} \label{hat_vi}
f(t, \hat{v})=0, \quad \text{if } |\hat{v}_i| \geq C t^{-2/\lambda^2}
\end{equation}
So using (\ref{f_vi}) and (\ref{hat_vi}) we get
\begin{equation*}
\rho (t) = \int_{|\hat{v}_i| \leq C t^{-2/\lambda^2}} f(t, \hat{v}) (1+|\hat{v}|^2)^{1/2} \,d\hat{v}
\end{equation*}
Note that since $f(t, \hat{v})$ is a constant along the characteristics,
\begin{equation}\label{f_f0}
|f(t, \hat{v})| \leq \|f_0\|:= \sup \{ |f(t_0, \hat{v})| : \text{ for all } \, \hat{v}\}
\end{equation}
So we obtain
\begin{align*}
\rho (t) &\leq C \int_{|\hat{v}_i| \leq C t^{-2/\lambda^2}} f(t, \hat{v}) \,d\hat{v}\\
		 &\leq C \|f_0\| t^{-6/\lambda^2}
\end{align*}
Also by (\ref{f_vi}) -- (\ref{f_f0}) we have
\begin{align*}
J_i (t) &= \int_{|\hat{v}_i| \leq C t^{-2/\lambda^2}}  f(t, \hat{v}) \hat{v}_i \,d\hat{v}\\
	 	 &\leq C t^{-2/\lambda^2}\int_{|\hat{v}_i| \leq C t^{-2/\lambda^2}} f(t, \hat{v}) \,d\hat{v}\\
		 &\leq C \|f_0\| t^{-8/\lambda^2}
\end{align*}
Similarly
\begin{align*}
S_{ij} (t) &= \int_{|\hat{v}_i| \leq C t^{-2/\lambda^2}} f(t, \hat{v}) \hat{v}_i \hat{v}_j (1+|\hat{v}|^2)^{-1/2}  \,d\hat{v}\\
	 	 &\leq C t^{-4/\lambda^2} \int_{|\hat{v}_i| \leq C t^{-2/\lambda^2}} f(t, \hat{v}) \,d\hat{v}\\
		 &\leq C \|f_0\| t^{-10/\lambda^2}
\end{align*}
Now let us estimate the ratios $J_i/ \rho$ and $S_{ij}/\rho$.
By means of (\ref{f_vi}) and (\ref{hat_vi}),
we get
\begin{align*}
\frac{J_i(t)}{\rho(t)} 
&= \frac{\int f(t, \hat{v})  \hat{v}_i \,d\hat{v}}{\int f(t, \hat{v}) (1+|\hat{v}|^2)^{1/2} \,d\hat{v}}\\
&\leq C t^{-2/\lambda^2} \frac{\int f(t, \hat{v}) \, d\hat{v}}{\int f(t, \hat{v}) (1+|\hat{v}|^2)^{1/2} \,d\hat{v}}\\
&\leq C t^{-2/\lambda^2}
\end{align*}
Similarly
\begin{align*}
\frac{S_{ij}(t)}{\rho(t)} 
&= \frac{\int f(t, \hat{v}) \hat{v}_i \hat{v}_j (1+|\hat{v}|^2)^{-1/2} \,d\hat{v}}{\int f(t, \hat{v}) (1+|\hat{v}|^2)^{1/2} \,d\hat{v}}\\
&\leq C t^{-4/\lambda^2}
\end{align*}
\EoPr
 
In this proposition since all components of the energy momentum tensor in an orthonormal frame
go to zero as $t$ goes to infinity,
it can be concluded that in a certain sense solutions of Einstein-Vlasov system 
coupled to a nonlinear scalar field with a exponential potential
are approximated by vacuum Einstein solutions.
In a more detailed level (\ref{ratio1}) and (\ref{ratio2}) resemble the non-tilted dust-like solutions 
in which $J_i(t)$ and $S_{ij}(t)$ are identically zero.

\section*{Acknowlegements}
The author thanks Alan~D.~Rendall for discussions on the subject of this paper.


\end{document}